\documentclass[12pt]{article}
\usepackage[dvips]{graphicx}
\begin{document}
\setcounter{section}{0}
\title{Time Operator for the Quantum Harmonic Oscillator:
Resolution of an Apparent Paradox}

\author{Alex Granik\thanks {Physics Department,University of the
                           Pacific,Stockton,CA. 95211}\hspace{2mm} and
H.Ralph Lewis\thanks{Dartmouth College,Dept.of Physics
                            and Astronomy,NH 03755}}
\maketitle

\begin{abstract}
An apparent paradox is resolved that concerns the existence of
time operators which have been derived for the quantum harmonic
oscillator. There is an apparent paradox because, although a time
operator is canonically conjugate to the Hamiltonian, it has been
asserted that no operator exists that is canonically conjugate to
the Hamiltonian. In order to resolve the apparent paradox, we
work in a representation where the phase operator is diagonal.
The boundary condition on wave functions is such that they be
periodic in the phase variable, which is related to the
(continuous) eigenvalue of the time operator. Matrix elements of
the commutator of the time operator with the Hamiltonian involve
the phase variable itself in addition to periodic functions of the
phase variable. The Hamiltonian is not hermitian when operating
in space that includes the phase variable itself. The apparent
paradox is resolved when this non-hermeticity is taken into
account correctly in the evaluation of matrix elements of the
commutation relation.
\end{abstract}
\pagebreak

\section{ Introduction}

In the space spanned by the eigenstates of the coordinates
operator $q$ and  the momentum operator $p$, we consider the
quantum harmonic oscillator described by the following Hamiltonian
\begin{equation}
\label{eq:to1}
 H= {\frac{1}{2m}}p^2+ \frac{1}{2}m\omega^2q^2
\end{equation}
and we consider a time operator $\chi$, which is an operator
conjugate to the Hamiltonian in that space:
\begin{equation}
\label{eq:to2}
 [\chi,H]= \it i\hbar
\end{equation}
where $\chi =\chi(q,p)$ is a function of the operators$(q,p)$,
but is not an explicit function of the time $t$. For an
eigenstate of $H$, the( continuous) eigenvalue of $\chi$ is an
angle variable whose period is energy-dependent and whose domain
is $(-\infty,\infty)$. It is convenient to introduce a phase
operator $\phi$ by \cite{hl1}
\begin{equation}
\label{eq:to3}
 \omega\chi = \frac{\pi}{2}\it 1 - \phi +G(H)
\end{equation}
where $G(H)$ is some function of $H$. For an eigenstate of $H$,
the (continuous) eigenvalue of $\phi$, which we denote by
$\varphi$ and call the phase variable, is also an angle whose
domain is $(-\infty,\infty)$; but its period is the constant
$2\pi$.

\smallskip
Aharanov and Bohm \cite{aa1} discussed the time operator for the
case of a free particle. A time operator for the harmonic
oscillator was derived by Bender and Dunne \cite{bb1}-\cite{bb3}
and was re-derived  using similar considerations by Lewis,
Lawrence, and Harris \cite{hl1}, who were motivated by its
application as a starting point of a perturbation theory for the
construction of invariant quantum operators \cite {hl2}

\smallskip
Susskind and Glogower \cite {ls1} showed that a time operator for
the harmonic oscillator does not exist in the Hilbert space of
energy eigenstates, and they proposed a pair of alternate
operators for use in this Hilbert space. Cannata and Ferrari
\cite{fc1,fc2} also proposed an alternate operator fro use in the
Hilbert space. In a Comment on the paper by Lewis {\it et al},
Smith and Vaccaro \cite{ts1} disputed the existence of a time
operator for the harmonic oscillator in any space by pointing at
an apparent inconsistency. They incorrectly asserted that,
because the eigenvalue spectrum of $H$ is discrete, the
commutation relation (\ref{eq:to2}) does not have a
representation in the Hilbert space of the eigenstates of $H$.
Although Lewis {\it et al} denied the validity of this apparent
inconsistency in a Reply \cite{hl3}, they did not address the
issue in detail. The existence of time operators  that have been
derived , together with the apparent inconsistency , constitutes
an apparent paradox. This apparent inconsistency is addressed in
this letter. It is removed by correctly taking into account the
non-hermeticity of the Hamiltonian in a space that contains the
phase variable itself in addition to periodic functions of the
phase variable. As a result, the commutation relation
(\ref{eq:to2}) is seen to have a representation in the Hilbert
space, and the apparent paradox is thus resolved. The present
authors believe that a time operator for the harmonic oscillator
can also be derived simply and directly from the classical time
function by adapting an Moyal quantization \cite{em1}. This
derivation is left for the future publication.

\smallskip
The apparent inconsistency raised by Smith and Vaccaro{ts1} is
based on the assertion that the relation
\begin{equation}
\label{eq:to4}
  <m| [B,H] |n> = <m|B|n>(n-m)\hbar\omega
\end{equation}
holds for any operator $B$, where
\begin{equation}
\label{eq:to5}
 H|n> = (n+\frac{1}{2})\hbar\omega|n>
 \end{equation}
 Relation (\ref{eq:to4}) is correct if operation by $B$ on any
 eigenstate of $H$ generates a state that lies in the space
 spanned by the eigenstates of $h$. However $B$ does not have that
 property if it is conjugate to $H$, as in the case with the time
 operator $\chi$. Then, relation (\ref{eq:to4}) is incorrect. This
 can be understood clearly by working in the phase variable
 representation, {\it i.e.,} in the representation of the
 eigenstates of $\varphi$; in that representation both $\varphi$
 and $\chi$ are diagonal. The action of $\varphi$ on one of its
 eigenstates is simply multiplication by the eigenvalue; and
 according to (\ref{eq:to3}), the action of $\chi$ on an eigenstate
 of $\varphi$ is multiplication by a linear function of $\varphi$.
 Both $\varphi$ and $\chi$ are hermitian in the full space of the
 eigenstates of $\varphi$ and in every  subspace thereof. The reason
 why (\ref{eq:to4}) is incorrect when applied to $<m| [\chi,H]
 |n>$ is that $H$ is {\it not hermitian} in the full space of the
 eigenstates of $\varphi$. The physical boundary condition on wave
 functions is that they be periodic functions of $\varphi$. In a
 space of periodic functions of $\varphi$, the Hamiltonian $H$ is
 hermitian. However in a space that includes aperiodic functions
 of $\varphi$, like variable $\varphi$ itself , $H$ is not
 hermitian. We evaluate  $<m| [\chi,H] |n>$ as follows
 \begin{equation}
 \label{eq:to6}
 \begin{array}{c}
 <m|[\chi,H]|n> = <m|\chi H|n>-<m|H\chi|n>\\
 =(n+\frac{1}{2})\hbar\omega <m|\chi|n>-<m|H\chi|n>,
 \end{array}
\end{equation}
where  $<m| [\chi,H]|n>$ has been evaluated using (\ref{eq:to5}).
If $H$ {\it were hermitian} then also the last term in
(\ref{eq:to6}) could be evaluated directly using (\ref{eq:to5}):
\begin{equation}
\label{eq:to7}
\begin{array}{c}
 <m|H\chi|n>=<n|\chi^{\dag} H^{\dag}|n>^*=(m+\frac{1}{2})\hbar\omega<n|\chi^{\dag}|m>^*
 =\\
(m+\frac{1}{2})\hbar\omega<m|\chi|n>
\end{array}
\end{equation}
This would give an evaluation of $<m|[\chi, H]|n>$ in agreement
with (\ref{eq:to4}). However, $H$ cannot be considered hermitian
in the evaluation of $<m|[\chi, H]|n>$ because the presence of
$\chi$ in $<m|[\chi, H]|n>$ leads to aperiodic functions of
$\varphi$. Therefore relation (\ref{eq:to4}) is invalid in this
case. The is consistent with the discussion in the introductory
section of a paper by Cannata and Ferrari \cite{fc2}, where it is
demonstrated that relation (\ref{eq:to4}) holds for any operator
$B$ whose action is confined to the space of eigenstates  of $H$,
{\it i.e.,} for any $B$ that is representable in the form $
B=\sum_{k,l}\beta_{kl}|k><l|$. Because the operators $\varphi$ and
$\chi$ introduce aperiodic functions of $\varphi$ ,they are not
representable in this form. A correct evaluation of $<m| [\chi,H]
|n>$ yields the result that corresponds to the canonical
commutation relation (\ref{eq:to2}). This is subject of Sec.2.

\section{Representation of the commutation relation}
We now proceed to evaluate the matrix elements of $<m|
[\chi,H]|n>$ directly in terms of the phase variable
representation. In terms of $\varphi$, the commutation relation
(\ref{eq:to2}) is
\begin{equation}
\label{eq:to8}
[\varphi,H]= -i\hbar\omega
\end{equation}
and the phase variable representation  of $H$ implied by
(\ref{eq:to8}) is
\begin{equation}
\label{eq:to9}
 H =i\hbar\omega \frac{d}{d\varphi} + f(\varphi)
\end{equation}
where $f(\varphi)$ is some function of $\varphi$. The normalized
periodic eigenfunctions of $H$ are
\begin{equation}
\label{eq:to10}
<\varphi|n>=\frac{1}{\sqrt{2\pi}}e^{-in\varphi}
\end{equation}
whose eigenvalues are $n\hbar\omega + f(\varphi)$. In order that
the eigenvalues of $H$ be correctly given as
$(n+\frac{1}{2}\hbar\omega)$, we must take $f(\varphi)
=\hbar\omega/2$. Thus $H$ is given
\begin{equation}
\label{eq:to11}
 H=i\hbar\omega\frac{d}{d\varphi}+ \frac{\hbar\omega}{2}
\end{equation}
We now express the matrix elements of the commutator as
\begin{equation}
\label{eq:to12}
\begin{array}{c}
 <m|[\chi,H]|n>=-\frac{1}{\omega}<m|[\phi,H]|n>\\
 =-(n+\frac{1}{2})\hbar<m|\phi|n>+ \frac{1}{\omega}<m|H\phi|n>
\end{array}
\end{equation}
and evaluate the last term as follows
\begin{equation}
\label{eq:to13}
\begin{array}{l}
 \frac{1}{\omega}<m|H\phi|n>=\int_{0}^{2\pi}<\varphi|m>^*
 H\varphi<\varphi|n>d\varphi\\
 =\frac{\hbar}{2}<m|\phi|n>+
 i\hbar\int_{0}^{2\pi}<\varphi|m>^*\frac{d}{d\varphi}<\varphi|n>d\varphi\\
 =\frac{\hbar}{2}<m|\phi|n>+i\hbar\int_{0}^{2\pi}\frac{d}{d\varphi}
 {<\varphi|m>^*\varphi<\varphi|n>}d\varphi\\
 -i\hbar\int_{0}^{2\pi}\varphi<\varphi|n>\frac{d}{d\varphi}
 <\varphi|m>^*d\varphi
 =i\hbar{<\varphi|m>^*\varphi<\varphi|n>}|_{0}^{2\pi}\\
 +(m+\frac{1}{2})\hbar<m|\phi|n>=i\hbar+(m+\frac{1}{2})\hbar<m|\phi|n>
 \end{array}
 \end{equation}
Using
\begin{equation}
\label{eq:to14}
\begin{array}{c}
<m|\phi|n>=\int_{0}^{2\pi}<\varphi|m>^*\varphi<\varphi|n>d\varphi
=-\frac{1}{m-n},\hspace{5mm} if \hspace{1mm} m\neq n \\
\hspace{7.7cm}=\hspace{3mm}0, \hspace{11mm}if\hspace{1mm} m=n
\end{array}
\end{equation}
we finally obtain for the matrix elements of the commutator
\begin{equation}
<m|[\chi,H]|n>=i\hbar+(m-n)\hbar<m|\phi|n>=i\hbar\delta_{mn}
\end{equation}
Thus, the commutation relation (\ref{eq:to2}) has a {\it correct}
representation in the space of eigenstates of $H$, despite of the
fact that $\chi$ {\it cannot} be represented in that space.
\pagebreak


\begin{thebibliography}{99}
\bibitem{hl1} H.R.Lewis,W.E.Lawrence,and J.D.Harris,
Phys.Rev.Lett.{\bf 77},5157(1996)
\bibitem{aa1} Y.Aharonov and D.Bohm, Phys.Rev. {\bf 122},1649
(1961)
\bibitem{bb1} C.M.Bender and G.V.Dunne, Phys.Rev.D {\bf 40},2739
(1989)
\bibitem{bb2} C.M.Bender and G.V.Dunne, Phys.Rev.D {\bf 40},3504
(1989)
\bibitem{bb3} C.M.Bender, Contemporary Math. {\bf 160},31-45
(1994)
\bibitem{hl2} H.R.Lewis,W.E.Lawrence, Phys.Rev. A {\bf55},2615
(1997)
\bibitem{ls1} L.Susskind and J.Glogower,Physics {\bf 1},49 (1964)
\bibitem{fc1} F.Cannata and L.Ferrari, Found,Phys.Lett. {\bf 4}
569 (1991)
\bibitem{fc2} F.Cannata and L.Ferrari, Found,Phys.Lett. {\bf 4}
557 (1991)
\bibitem{ts1} T.B.Smith and J.A.Vaccaro, Phys.Rev.Lett. {\bf 80},
2745 (1998)
\bibitem{hl3} H.R.Lewis,W.E.Lawrence,and J.D.Harris,
Phys.Rev.Lett.{\bf 80},2746 (1998)
\bibitem{em1}J.E.Moyal, Proc.Cambridge Phil.Soc. {\bf 45}, 99
(1949)
\end{thebibliography}
\end{document}